\documentclass[%
 reprint,
superscriptaddress,
 amsmath,amssymb,
 aps,
floatfix,
]{revtex4-2}

\usepackage{graphicx}
\usepackage{dcolumn}
\usepackage{bm}

\usepackage{pxfonts}
\usepackage{stfloats}
\usepackage{float}
\usepackage{siunitx}
\usepackage{hyperref}
\hypersetup{
  colorlinks   = true, 
  urlcolor     = blue, 
  linkcolor    = blue, 
  citecolor   = blue 
}
\usepackage{lipsum}
\usepackage{chngcntr}

\begin{document}

\preprint{APS/123-QED}

\title{Local Nonlinear Elastic Response of Extracellular Matrices}

\author{Haiqian Yang}
\affiliation{Department of Mechanical Engineering, Massachusetts Institute of Technology, Cambridge, MA 02139, USA}
\author{Estelle Berthier} 
\affiliation{Arnold-Sommerfeld-Center for Theoretical Physics and Center for NanoScience, Ludwig-Maximilians-Universität München, D-80333 München, Germany}
\author{Chenghai Li}
\affiliation{Department of Mechanical and Aerospace Engineering, University of California, San Diego, La Jolla, CA 92093, USA}
\author{Pierre Ronceray}
\affiliation{Aix Marseille Univ, CNRS, CINAM, Turing Center for Living Systems, Marseille, France}
\author{\\Yu Long Han}
\affiliation{Department of Mechanical Engineering, Massachusetts Institute of Technology, Cambridge, MA 02139, USA}
\author{Chase P. Broedersz}
\affiliation{Arnold-Sommerfeld-Center for Theoretical Physics and Center for NanoScience, Ludwig-Maximilians-Universität München, D-80333 München, Germany}
\affiliation{Vrije Universiteit Amsterdam, Department of Physics and Astronomy, 1081 HV Amsterdam, The Netherlands}
\author{Shengqiang Cai}
\affiliation{Department of Mechanical and Aerospace Engineering, University of California, San Diego, La Jolla, CA 92093, USA}
\author{Ming Guo}\thanks{guom@mit.edu}
\affiliation{Department of Mechanical Engineering, Massachusetts Institute of Technology, Cambridge, MA 02139, USA}

\date{\today}

\begin{abstract}
Nonlinear stiffening is a ubiquitous property of major types of biopolymers that make up the extracellular matrices (ECM) including collagen, fibrin and basement membrane. Within the ECM, many types of cells such as fibroblasts and cancer cells are known to mechanically stretch their surroundings that locally stiffens the matrix. Although the bulk nonlinear elastic behaviors of these biopolymer networks are well studied, their local mechanical responses remain poorly characterized. Here, to understand how a living cell feels the nonlinear mechanical resistance from the ECM, we mimic the cell-applied local force using optical tweezers; we report that the local stiffening responses in highly nonlinear ECM are significantly weaker than responses found in bulk rheology, across two orders of magnitude of the locally applied force since the onset of stiffening. With a minimal model, we show that a local point force application can induce a stiffened region in the matrix, which expands with increasing magnitude of the point force. Furthermore, we show that this stiffened region behaves as an effective probe upon local loading. The local nonlinear elastic response can be attributed to the nonlinear growth of this effective probe that linearly deforms an increasing portion of the matrix.
\end{abstract}

\maketitle

\section{INTRODUCTION}
Nonlinear stiffening is ubiquitous in biopolymers~\cite{1-storm2005nonlinear,xu2015nonlinearities}, including the ECM polymers such as collagen, fibrin and basement membrane~\cite{piechocka2010structural,stein2011micromechanics,11-licup2015stress,munster2013strain,sharma2016strain,2-han2018cell,li2021nonlinear}. In these ECM materials, cells such as fibroblasts and certain types of cancer cells are known to deform and stiffen the matrix~\cite{2-han2018cell,3-hall2016fibrous,jansen2013cells,steinwachs2016three,burkel2017microbuckling,krajina2021microrheology,van2016strain}. The mechanical nonlinearity is a remarkable feature as it allows the cell to spontaneously modify their mechanical environments simply by pulling on the matrices. While the mechanical properties of the ECM are known to be important regulators of cell functions~\cite{4-discher2005tissue,5-engler2006matrix}, little is known about the interplay between cells and nonlinear elasticity of the ECM.

\begin{figure*}
\includegraphics[width=.8\textwidth]{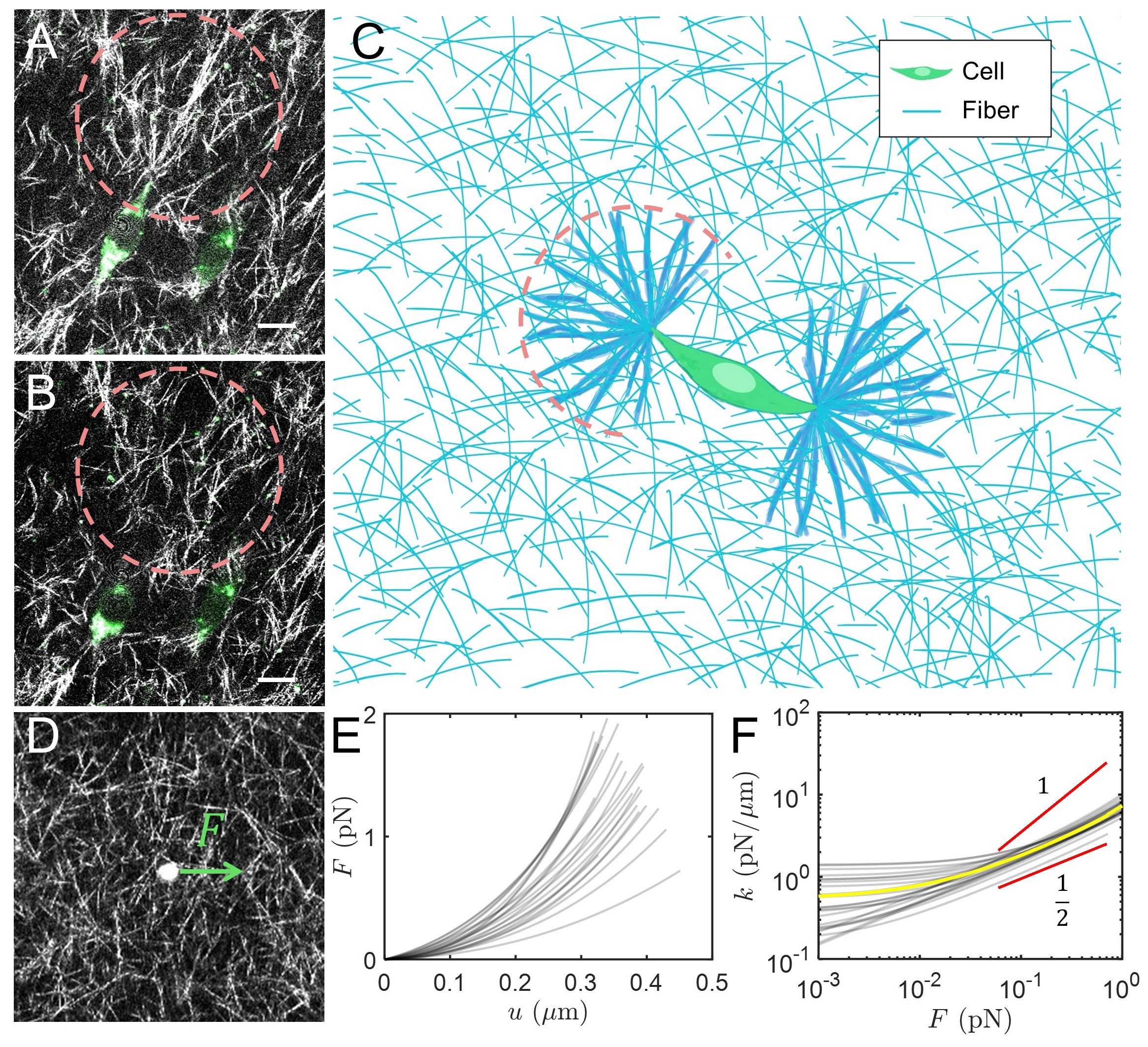}
\caption{The force applied by a single cell is highly localized, and local nonlinear stiffening of collagen gels deviates from the scaling law in bulk. (\textit{A}) Image of a mouse embryonic fibroblast (green, cytoplasm) within \SI{2}{\mg \ml\tothe{-1}} collagen networks (grey) stretching the matrix with filopodia. (\textit{B}) Image of the same cell with one filopodium released. The dash circle schematically marks the highly deformed region of the network, which is released once the force is removed. (scale bar, \SI{10}{\micro\meter}.) (\textit{C}) Schematics of a cell embedded in 3D fibrous ECM. The filopodium applies a point force to the matrix and creates a highly deformed zone. (\textit{D}) Confocal reflective image of a particle (\SI{2}{\micro\meter} in diameter) embedded in a \SI{4}{\mg \ml\tothe{-1}} collagen gel. A focused 1064-nm laser is used to trap the particle and apply a point force $F$. (\textit{E}) Force $F$ is a nonlinear function of displacement $u$. (\textit{F}) Starting from the onset of nonlinear stiffening, differential stiffness $k$ deviates from the bulk scaling law for two orders of magnitude (n=23). Yellow line is the average stiffness as a function of force.}
\label{fig:FIG1}
\end{figure*}

Existing studies on the nonlinear elastic properties of biopolymers primarily focus on the bulk response~\cite{6-mackintosh1995elasticity,7-broedersz2014modeling,8-broedersz2012filament,9-fung1967elasticity,10-fung2013biomechanics,11-licup2015stress,munster2013strain,sharma2016strain,13-ban2019strong,piechocka2010structural}. One important example is collagen, which is the most abundant constituent protein of the ECM~\cite{gelse2003collagens,stein2011micromechanics}. It was found that collagenous tissues can be well characterized by an exponential force-extension relation~\cite{9-fung1967elasticity,10-fung2013biomechanics}. Later on, it has been proven that the exponential stiffening behavior of collagen gels can be derived from the mechanics of fibrous networks~\cite{11-licup2015stress}. For collagen gels under shear stress $\sigma$ and shear strain $\gamma$, a direct consequence is that the differential stiffness $K ~(K=\frac{d\sigma}{d\gamma})$ is directly proportional to stress $\sigma$ in the nonlinear regime for bulk measurements, $K\sim\sigma$~\cite{11-licup2015stress}.

Cells embeded in the ECM, however, interact with the surrounding matrix locally instead of in bulk. For example, a fibroblast in the collagen matrix extends its filopodia and pulls on the fibers (Fig.~\ref{fig:FIG1}A-C). The loading condition from a cell is localized, and it generates a nonuniform deformation field in the matrix, which is distinct from a uniform bulk measurement~\cite{burkel2017mechanical,proestaki2019modulus}. It remains elusive how much mechanical resistance a cell feels when applying such localized forces. Within highly nonlinear elastic matrices, while it is still expected that a cell will face more resistance as it pulls harder, it is not clear if the resistant force grows exponentially with displacement such that the resistant stiffness $k$ grows linearly with the force $F$ as expected from bulk measurements. In local microrheolgy by optical and magnetic tweezers, where a particle of diameter $d$ is moved to a displacement $u$ by a force $F$, nominal stress $\sigma=\frac{4F}{\pi d^2}$ and nominal strain $\epsilon=u/d$ are usually calculated~\cite{chapman2014nonlinear,robertson2018optical}, which implies the local measurements are analogous to the bulk measurements~\cite{robertson2018optical}. While the linear responses can be quite similar~\cite{staunton2016mechanical}, our results indicate local and bulk nonlinear power-law responses are not necessarily consistent. In fact, to generate the same amount of displacement, our results suggest that the required localized force is much smaller than that expected from bulk mechanics. Mimicking this localized loading condition, here we show experimental results by optical tweezers that the nonlinear response of collagen to a local force deviates from the expectation from the bulk response $k \sim F$, but instead is closer to $k \sim F^{1/2}$. We then explain this weaker power-law relation by a minimal theory of an effective probe: the observed local nonlinear response is similar to the case of using a larger effective probe to linearly deform the far-field matrix whose shear modulus is $G$, and the effective probe size $R^*$ grows as ${(\frac{F}{G})}^{1/2}$. Surprisingly, this argument suggests $k \sim F^{1/2}$ is a universal relation for local mechanical measurements independent of bulk stiffening behaviors provided that the stiffening is sufficiently strong. Our results highlight the significant differences between nonlinear responses upon local and bulk force applications, which is highly relevant in cell-ECM interactions and thus mechanobiology of multicellular living systems.

\section{MATERIALS AND METHODS}
\subsection*{Preparation of Collagen Gels}
To make \SI{1}{\ml} \SI{4}{\mg \ml\tothe{-1}} collagen gels, \SI{400}{\micro\liter} \SI{10}{\mg \ml\tothe{-1}} collagen solution (Advance Biomatrix, Cat\# 5133) is mixed with \SI{460}{\micro\liter} Dulbecco's Modified Eagle Medium and \SI{100}{\micro\liter} $10\times$ phosphate-buffered saline. \SI{40}{\micro\liter} 0.1 M NaOH is then added to the mixture to adjust pH to 7.2-7.6. The solution is then polymerized in the cell culture incubator (\SI{37}{\degreeCelsius}, 5\% $\mathrm{CO_2}$) for 60 minutes.
\subsection*{Preparation of Fibrin Gels}
To make \SI{1}{\ml} \SI{3}{\mg \ml\tothe{-1}} fibrin gels, \SI{500}{\micro\liter} \SI{6}{\mg \ml\tothe{-1}} fibrinogen (F8630-1G, Sigma-Aldrich) solution is mixed with \SI{500}{\micro\liter} \SI{4}{U~\ml\tothe{-1}} thrombin (T4648-1KU, Sigma-Aldrich). The solution is then polymerized in the cell culture incubator (\SI{37}{\degreeCelsius}, 5\% $\mathrm{CO_2}$) for 20 minutes.

\subsection*{Optical Tweezers Measurements}
Microparticles (\SI{2}{\micro\meter}, C37278, ThermoFisher, and \SI{4.5}{\micro\meter} carboxylate microspheres, Polyscience) are embeded in the gel prior to polymerization. Home-built optical tweezers are used~\cite{gupta2020quantification}. The laser beam is moved at a constant velocity of \SI{1}{\micro\meter \s\tothe{-1}}. The relative distance between the laser and the bead is recorded. Raw data is processes as described in~\cite{2-han2018cell}. The tweezers is calibrated in phosphate-buffered saline solution. The Power Spectrum Density method is used ~\cite{gupta2020quantification}. The trap stiffness is \SI{15}{\pico\newton/\micro\meter}.

\subsection*{Finite Element Simulations}
For local measurement, an axial symmetric geometry is used, where a rigid spherical domain of radius $R_0$ is embeded in an cylindrical matrix of radius $R=250R_0$ and height $H=2R$. The matrix is treated as incomprresible, and various forms of hyperelastic constitutive relations are assigned to the matrix. The lateral surface of the matrix is fixed and the top and bottom surfaces are traction free. A series of point forces are applied at the center of the rigid spherical domain and the displacement is recorded. For the bulk measurement, a rectangular domain with the same material properties is sheared by a shear stress $\sigma$ to a shear strain $\gamma$, and the differential modulus can be defined as $K=d\sigma/d\gamma$. COMSOL Multiphysics is used for the simulation.

\subsection*{Network Simulations}
Details of the network simulation are in~\cite{Estelle}. Briefly, a point-force monopole is placed at the center of a 3D disordered network of radius $R=40$. The 3D network is generated by placing fibers on a face-centered cubic lattice and randomly depleting it such that a bond is present with probability $p=0.4$. Fibers are discretized with bonds that act as springs with linear stretching modulus $\mu = 100$ and nonlinear longitudinal response $exp(\mu\delta\varepsilon)-1$, where $\delta \varepsilon$ is the bond deformation. The fibers also resist transverse deflection with a bending rigidity $\kappa = 1$. Fibers are connected by freely deforming hinges at their intersection. Results are averaged over 100 random network realizations.

\begin{figure*}
\includegraphics[width=.8\textwidth]{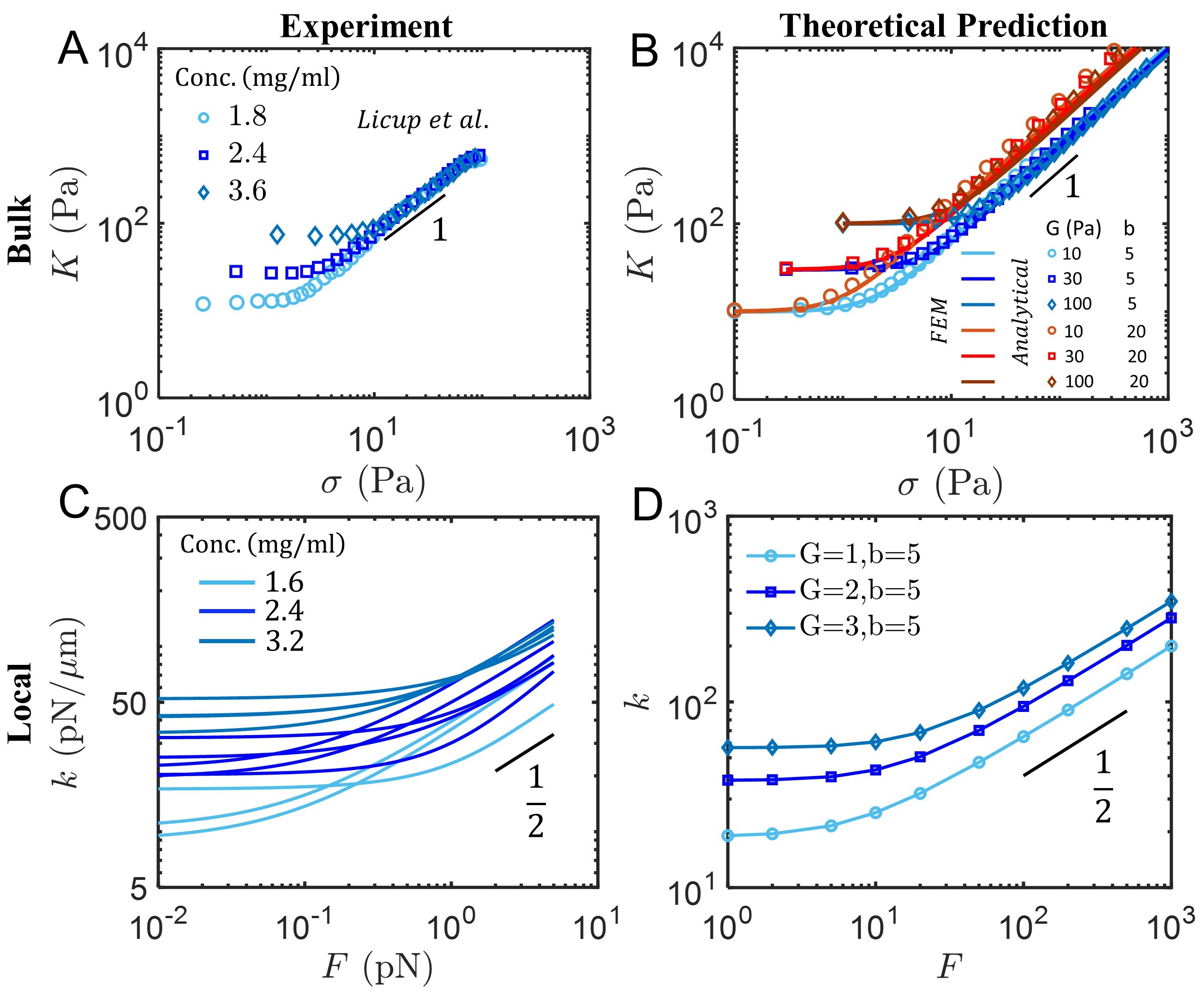}
\caption{A minimal model predicts both bulk and local stiffening behaviors of collagen gels. Bulk stiffening of collagen follows $K \sim \sigma$ in both experiment (\textit{A}, figure reproduced from Licup et al.~\cite{11-licup2015stress}), and theory (\textit{B}). (\textit{C}) Unidirectional forces are applied on particles (\SI{4.5}{\micro\meter} in diameter) embeded in collagen gels at various concentrations. Local scaling law is close to $k \sim F^{1/2}$. (\textit{D}) The power-law relation $k \sim F^{1/2}$ is recovered by theory.}
\label{fig:FIG2}
\end{figure*}

\section{RESULTS AND DISCUSSION}
\subsection*{Local stiffening in collagen deviates from bulk stiffening}
To mimic the localized force exerted by filopodia, we use optical tweezers to displace micron-sized particles embedded in collagen gels. A bead is displaced to a distance $u$ from its equilibrium position by an optical force $F$ (Fig.~\ref{fig:FIG1}D), and the differential stiffness $k$ can be calculated by $k=\frac{dF}{dx}$. When we perform this measurement in \SI{4}{\mg \ml\tothe{-1}} collagen gels with \SI{2}{\micro\meter}-in-diameter beads, we find that the material indeed stiffens with increasing force (Fig.~\ref{fig:FIG1}E); the differential stiffness $k$ shows a linear regime followed by a nonlinear regime (Fig.~\ref{fig:FIG1}F). Interestingly, the logarithmic slope in the nonlinear regime clearly deviates from 1 since the initiation of nonlinear stiffening and ranging 2 orders of magnitude of the applied force ($10^{-2} \sim 1$ pN), in contrast to bulk measurements where stiffness reaches the bulk scaling law $K \sim \sigma$ rather quickly within one order of magnitude of applied stress (Fig.~\ref{fig:FIG2}A). A closer examination reveals that the logarithmic slope is about $1/2$, suggesting $k \sim F^{1/2}$ (Fig.~\ref{fig:FIG1}F); this is consistent with directly inspecting the force $F$ as a function of the displacement $u$, that the logarithmic slope changes from 1 to 2 (Fig.~\ref{fig:FIGS1}), indicating the relation between force $F$ and displacement $u$ changes from a linear function $F\sim u$ to a quadratic function $F\sim u^2$. This result indicates that the displacement induced by a localized force is greater than the expectation from the bulk mechanics. To further confirm this behavior, we perform this experiment in collagen gels of different concentrations (e.g. 1.6, 2.4 and 3.2 \SI{}{\mg \ml\tothe{-1}}) with a different particle size (\SI{4.5}{\micro\meter} in diameter); these experiments all show logarithmic slopes close to $1/2$ in the nonlinear regime (Fig.~\ref{fig:FIG2}C). Moreover, we perform measurements on fibrin gels. Remarkably, the local nonlinear response of fibrin gels is also close to $k \sim F^{1/2}$ (Fig.~\ref{fig:FIGS2} and Fig.~\ref{fig:FIG3}), regardless of the different bulk stiffening response of collagen and fibrin gels~\cite{piechocka2010structural,11-licup2015stress}. These results demonstrate the generality of this weak power-law relation $k \sim F^{1/2}$ in local micromechanics measurements.

\subsection*{Theoretical predictions of both bulk and local power-law responses}
To understand this weaker stiffening behavior $k \sim F^{1/2}$ in local measurements, we consider a minimal material model of an incompressible Fung-type elastic medium with an elastic free energy function $W_{exp}=\frac{G}{2b} \exp (b(I_1-3))$, where $G$ is the linear shear modulus, $b$ is a stiffening parameter, and $I_1$ is the first invariant of the deviatoric Cauchy-Green deformation tensor. As expected, the Fung-type material relation can capture the exact responses in bulk measured by rheometers (Fig.~\ref{fig:FIG2}A\&B)~\cite{11-licup2015stress}. Further, we study the local response by performing finite element analysis of an axial symmetric model where a rigid particle of radius $R_0$ is embedded in a cylindrical matrix with radius $R=250R_0$ and height $H=2R$. The particle is perturbed in the axial direction from its equilibrium position by a force $F$ to a distance $u$. Interestingly, we find that this minimal model can recover the $1/2$ scaling law observed in our microrheology experiments of the local response (Fig.~\ref{fig:FIG1}F\&~\ref{fig:FIG2}C). It is worth noting that this model primarily focuses on the nonlinear elastic property, and other complex aspects such as heterogeneity, anisotropy, viscoelasticity and plasticity~\cite{beroz2017physical,jones2015micromechanics,nam2016strain,munster2013strain,kim2017stress} are not considered, such that the only difference between local and bulk measurements is geometry. This suggests that the difference in geometry alone may be the reason for this weaker stiffening behavior measured at the local scale. Interestingly, in a smaller matrix, the model predicts that this local response eventually becomes a global shear with increasing force, then the bulk response $k \sim F$ is recovered~(Fig.~\ref{fig:FIGS3}). However, due to limitations in the maximum force, this finite size effect is not clearly observed in our optical-tweezers experiments.

\begin{figure}[ht]
\centering
\includegraphics[width=.45\textwidth]{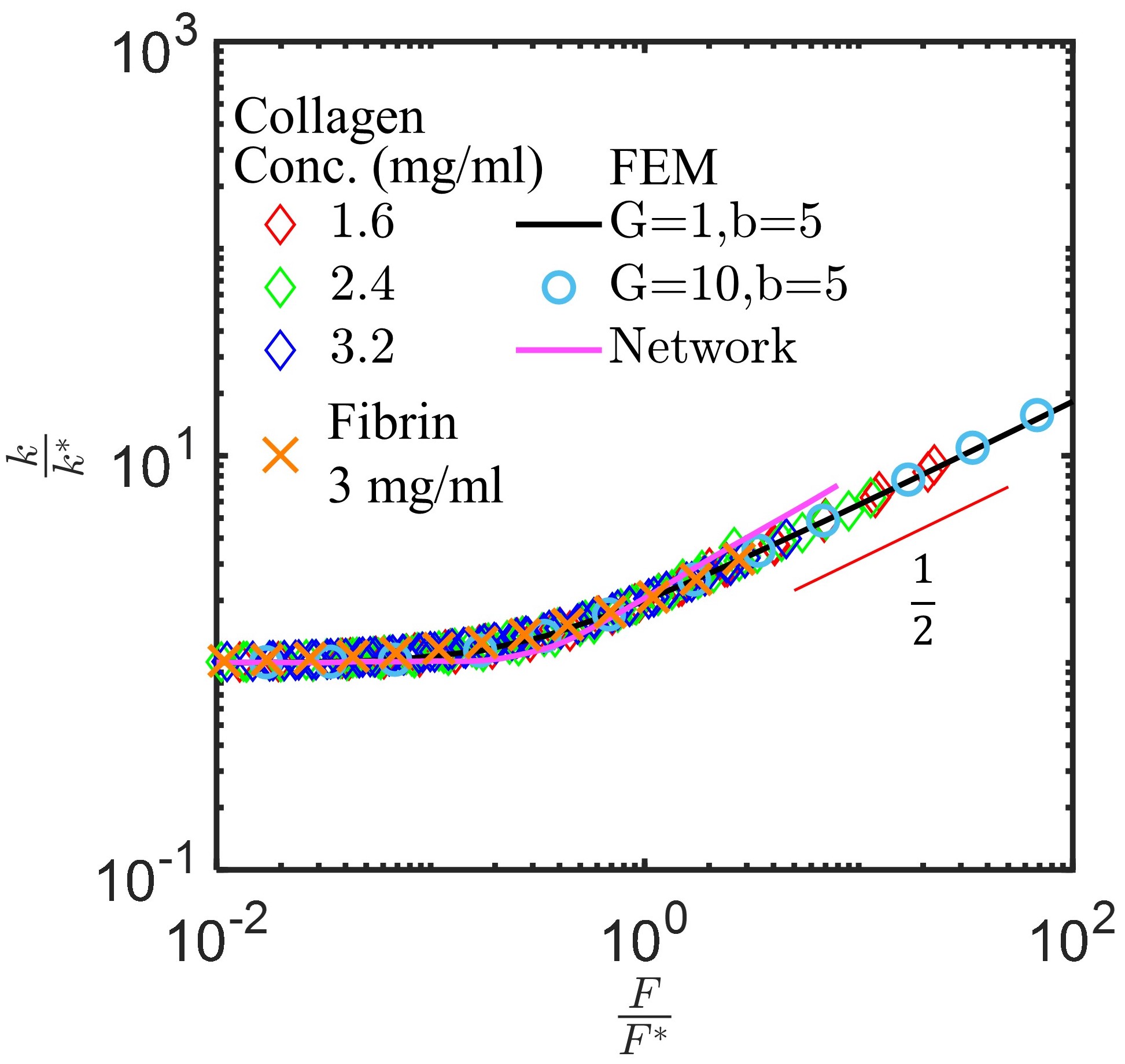}
\caption{All the experimental and theoretical curves can be collapsed to a single master curve by nondimensionalizing stiffness and force. The master curve shows $k \sim F^{1/2}$ in the nonlinear regime. The characteristic force $F^*$ is defined as the force at which the differential stiffness is twice the linear stiffness.}
\label{fig:FIG3}
\end{figure}

To further validate this theoretical prediction, we also perform fiber network simulations~\cite{2-han2018cell,Estelle}, where a point force $F$ is applied at the center of a spherical network. A similar power-law relation of the average measured stiffness as a function of the applied force is observed in the network simulations~(Fig.~\ref{fig:FIG3} \& Fig.~\ref{fig:FIGS4}B)~\cite{Estelle}. When we rescale the curves by defining the unidimensional stiffness $k/k^*$ and unidimensional force $F/F^*$ with linear stiffness $k^*$ and characteristic force $F^*$ at the onset of nonlinear stiffening, all curves collapse onto a single master curve, demonstrating the robustness of the $k \sim F^{1/2}$ scaling law (Fig.~\ref{fig:FIG3}). This result also suggests that both the discrete and continuous models capture the underlying physics of the experimental measurements.

\subsection*{Increase of an effective probe size}

\begin{figure*}[ht]
\centering
\includegraphics[width=.8\textwidth]{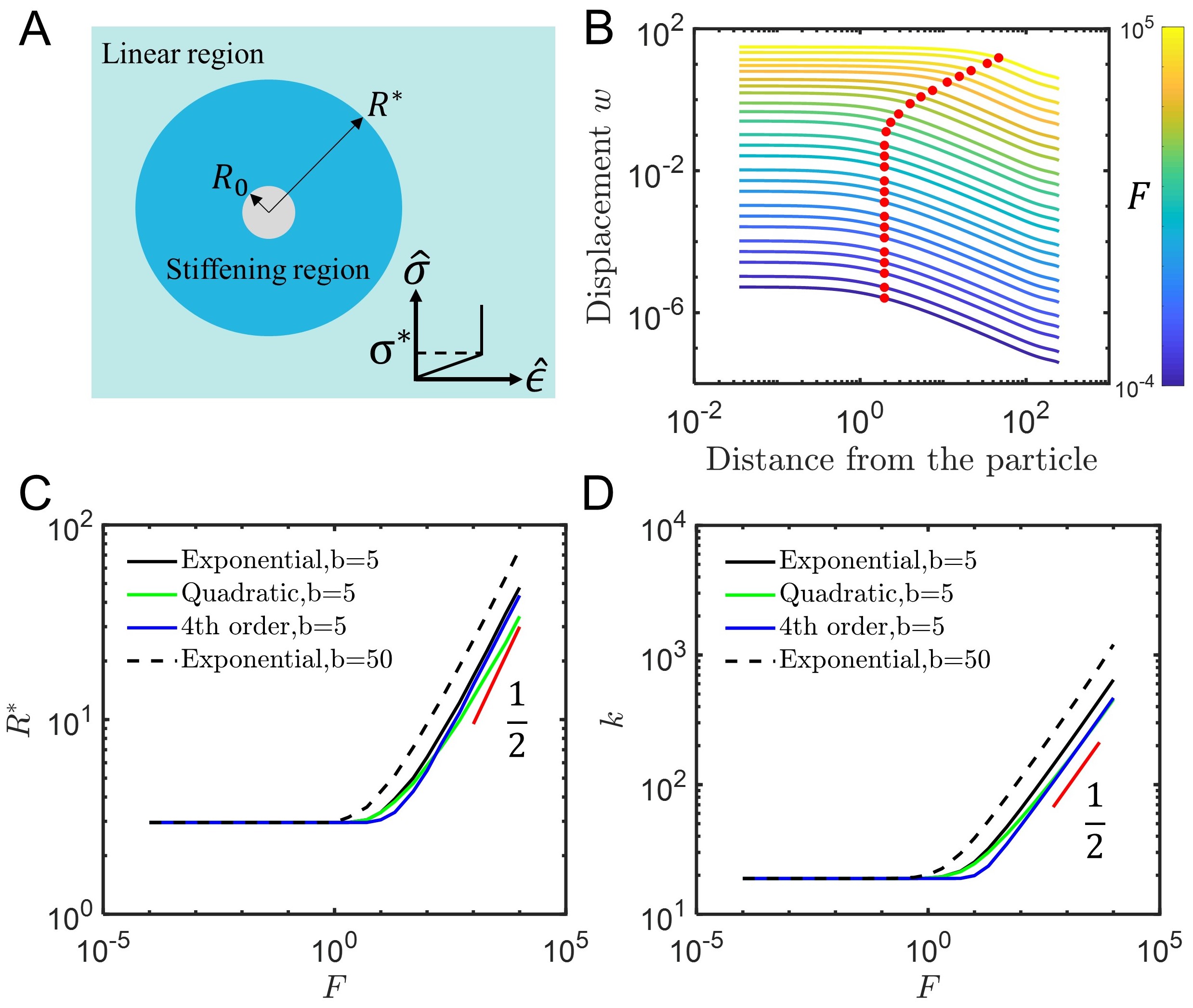}
\caption{The local stiffening response can be attributed to the increase of an effective probe size $R^*$, almost independent of the constitutive relations. (\textit{A}) The immediate adjacency of the particle is stiffened over the length scale $R^*$ such that the point force is probing the far-field linear matrix with an effective larger probe of length scale $R^*$. (Inset) The extreme case is an 'elastic-rigid' simplification that the material is linear at small stresses but becomes rigid as the stress grows larger than a threshold $\sigma^*$. (\textit{B}) $R^*$ can be estimated by the transition point of the axial displacement profile near the particle. $R^*$ is estimated as the distance when the axial displacement decays to half of that of the particle. The transition points are marked by red dots. $(G=1, b=5)$. (\textit{C}) Effective probe size $R^*$ grows as $F^{1/2}$ regardless of the constitutive relations. (\textit{D}) Stiffness $k$ grows as $F^{1/2}$ regardless of the constitutive relations. $(G=1)$.}
\label{fig:FIG4}
\end{figure*}

To understand this $1/2$ power-law relation, we first follow a dimensional analysis of the same boundary value problem as defined in the minimum finite element simulation. Considering the force balance, the point force to perturb the particle is balanced by the integral of traction fields on an arbitrary closed surface containing this point force, $F=-\oiint\sigma_{ij} n_j e_i^1 dA$, where $\sigma_{ij}$ are the stress components, $n_j$ are components of the surface normal vector and $e_i^1$ are components of the unit vector along the point force. In the matrix, strain and stress fields decay away from the probe. Although $F$ can be comparably large that nonlinearly deforms the vicinity of the probe, because the strain and stress fields approach 0 at infinity, there must be a closed surface above which the material only undergoes moderate strain, and thus behaves almost linearly. A stress scale on this surface is the linear shear modulus $G$. Denoting the length scale of this closed surface as $R^*$, the force balance equation yields $F \sim G R^{*2}$, so $R^*$ scales as $(\frac{F}{G})^{1/2}$. Interestingly, although fiber buckling is not explicitly considered in this model, the scaling law $R^* \sim F^{1/2}$ also coincides with the range of the buckled fibers in discrete fibrous models~\cite{12-ronceray2016fiber,Estelle}. Because the material within this range $R^*$ is more severely deformed and much stiffer than the outside as a result of nonlinear elastic stiffening, effectively we are using a larger rigid particle of length scale $R^*$ to probe the linear matrix surrounding this rigid particle. This yields $k \sim GR^*$ that is similar to the Stokes-Einstein law~\cite{14-squires2010fluid}, which we further verify in fiber network simulations by linearly probing the networks with particles of different sizes (Fig.~\ref{fig:FIGS4}A). Consequently, the stiffness $k \sim R^* \sim F^{1/2}$ (Fig.~\ref{fig:FIG4}D). As we examine the profile of displacement in the direction of the applied force in simulation, the direct adjacency of the particle moves with it almost rigidly; indeed, there is a critical length $R^*$ above which the displacement field starts to rapidly decay (Fig.~\ref{fig:FIG4}B). By perturbing the particle at different forces, we find $R^* \sim F^{1/2}$ (Fig.~\ref{fig:FIG4}C). This argument is similar to a boundary layer analysis in fluid mechanics, where a length scale $\delta$ can be derived by balancing viscous force with kinetic force; while the flow inside the boundary layer is highly viscous, the flow outside the boundary layer can be treated as inviscid flow. 

This dimensional argument appears to be independent of how the material is stiffened, and the only requirement is that the material stiffens strongly enough such that the material within $R^*$ can be considered as ‘rigid’. One extreme case is thus an ‘elastic-rigid material’ simplification: the material is linear at small stresses but becomes rigid once the stress exceeds a critical value $\sigma ^*$ (Fig.~\ref{fig:FIG4}A,~inset). For a perfect elastic-rigid material subjected to a point force $F$, $F \sim \sigma ^* R^{*2}$ is expected at the boundary of the rigid zone, such that $R^* \sim (\frac{F}{\sigma^*})^{1/2}$. The remaining issue is how strong the stiffening is to be approximated as an ‘elastic-rigid’ model. To investigate this, we further modify the simulation with different types of stiffening: the quadratic energy function is defined as $W_2=\frac{G}{2} (I_1-3)+ \frac{Gb}{4} (I_1-3)^2$, and the fourth order energy function is $W_4=\frac{G}{2} (I_1-3)+\frac{Gb}{8} (I_1-3)^4$. Under simple shear $\gamma$ and at the asymptotic limits of large $\gamma$, the energy scales as $\gamma^4$ and $\gamma^8$ respectively, thus the stiffness in bulk scales as $\sigma^{2/3}$ and $\sigma^{6/7}$. Interestingly, except for the quadratic stiffening that is not sufficiently strong and thus slightly deviates from $F^{1/2}$, the local stiffening behavior $k \sim F^{1/2}$ seems quite robust, regardless of the different types of stiffening in bulk (Fig.~\ref{fig:FIG4}C\&D).

The solution of $R^*$ may be written as $R^*=C(\frac{F}{G})^{1/2}$ where the prefactor $C$ cannot be derived from scaling analysis but should come from solving the boundary value problem either analytically or numerically. The prefactor $C$ should depend on the nonlinear dimensionless parameters. This minimal model suggests that for a material whose nonlinear term grows sufficiently fast, the nonlinear differential stiffness $k$ probed by the particle mainly comes from the response to an effective probe with size $R^*$ that nonlinearly grows with the force $F$ by $R^*\sim F^{1/2}$; this effective probe deforms a larger portion of the matrix with increasing force $F$, that yields $k \sim F^{1/2}$. Therefore, the nonlinear local response is a consequence of both material nonlinearity and this special geometry. Once the materials surrounding the probe becomes stiff enough compared to the linear far-field matrix, the response becomes dominated by the effective probe (Fig.~\ref{fig:FIG4}C\&D). Nevertheless, the onset of stiffening still indicates material nonlinearity; materials with stronger nonlinearity will stiffen at smaller forces (Fig.~\ref{fig:FIG4}C, exponential energy with $b=5$ and $b=50$). Indeed, the stiffening parameter $b$ determines the stiffening onset nominal strain $\epsilon^* = \frac{2u^*}{d}$, thus in principal the onset strain can be used to measure the stiffening factor of the material (Fig.~\ref{fig:FIGS5}).

\section{CONCLUSION}
We report the $K \sim \sigma$ relationship observed in exponentially stiffening material in bulk no longer holds for localized force applications. Instead, we find the relation is closer to $k \sim F^{1/2}$ (Fig.~\ref{fig:FIG1}F). Our results suggest that localized forces such as those generated by filopodia face less resistance and can generate larger displacements. This weaker power-law behavior is robust as can be observed in various types of materials, both in experiments and simulations (Fig.~\ref{fig:FIG3}). Furthermore, we show that the local response ($k \sim F^{1/2}$) can be explained by introducing an effective probe that grows with increasing forces (Fig.~\ref{fig:FIG4}).

Our results highlight the difference of highly nonlinear ECM gels between responses to local and bulk forces with simple concepts. While the mechanical fields are rather uniform under bulk forces, they decay away under localized forces. We show the local stiffening behaviors can be explained by considering the mechanical fields as two successive regions, i.e. an effective probe induced by material nonlinear stiffening and a linear matrix that is being deformed by the effective probe (Fig.~\ref{fig:FIG4}A). The nonlinear response to local forces ($k \sim F^{1/2}$) thus comes from this unique geometry that a growing effective probe is deforming an increasing portion of the far-field matrix.

We therefore propose a hypothesis that cells embeded in 3D strongly nonlinear matrices do not directly sense the material nonlinear elasticity, but instead through an effective probe that nonlinearly grows. Facilitated by the nonlinear material properties, the localized force exerted by a cell can create a stiff effective probe whose radius increases nonlinearly with the applied force $R^* \sim F^{1/2}$ through which a cell can deform farther matrices. This extended matrix deformation can facilitate mechanical communications between cells. These results may provide insights into understanding cell-matrix interactions, and mechanobiology of complex living systems such as tissues and organs.

\begin{acknowledgments}
We would like to acknowledge the support from the NIH (1R01GM140108), the MathWorks, and the Jeptha H. and Emily V. Wade Award at the Massachusetts Institute of Technology. M.G. acknowledges the Sloan Research Fellowship. E.B. and C.P.B.: This project has received funding from the European Union’s Horizon 2020 research and innovation programme under the Marie Sklodowska-Curie grant agreement No 891217 and the Deutsche Forschungsgemeinschaft (DFG, German Research Foundation) - Project-ID 201269156 - SFB 1032 (Project B12). P.R. is supported by France 2030, the French National Research Agency (ANR-16-CONV-0001) and the Excellence Initiative of Aix-Marseille University - A*MIDEX.
\end{acknowledgments}

\bibliography{apssamp}

\clearpage
\onecolumngrid
\LARGE \textbf{Supplementary Information for}
\vspace{1em}

\Large Local Nonlinear Elastic Response of Extracellular Matrices
\vspace{1em}

\small Haiqian Yang, Estelle Berthier, Chenghai Li, Pierre Ronceray, Yu Long Han, Chase P. Broedersz, Shengqiang Cai, Ming Guo
\vspace{10em}

\setcounter{figure}{0}
\renewcommand{\figurename}{FIG.}
\renewcommand{\thefigure}{S\arabic{figure}}
\begin{figure}[H]
\centering
\includegraphics[width=0.48\textwidth]{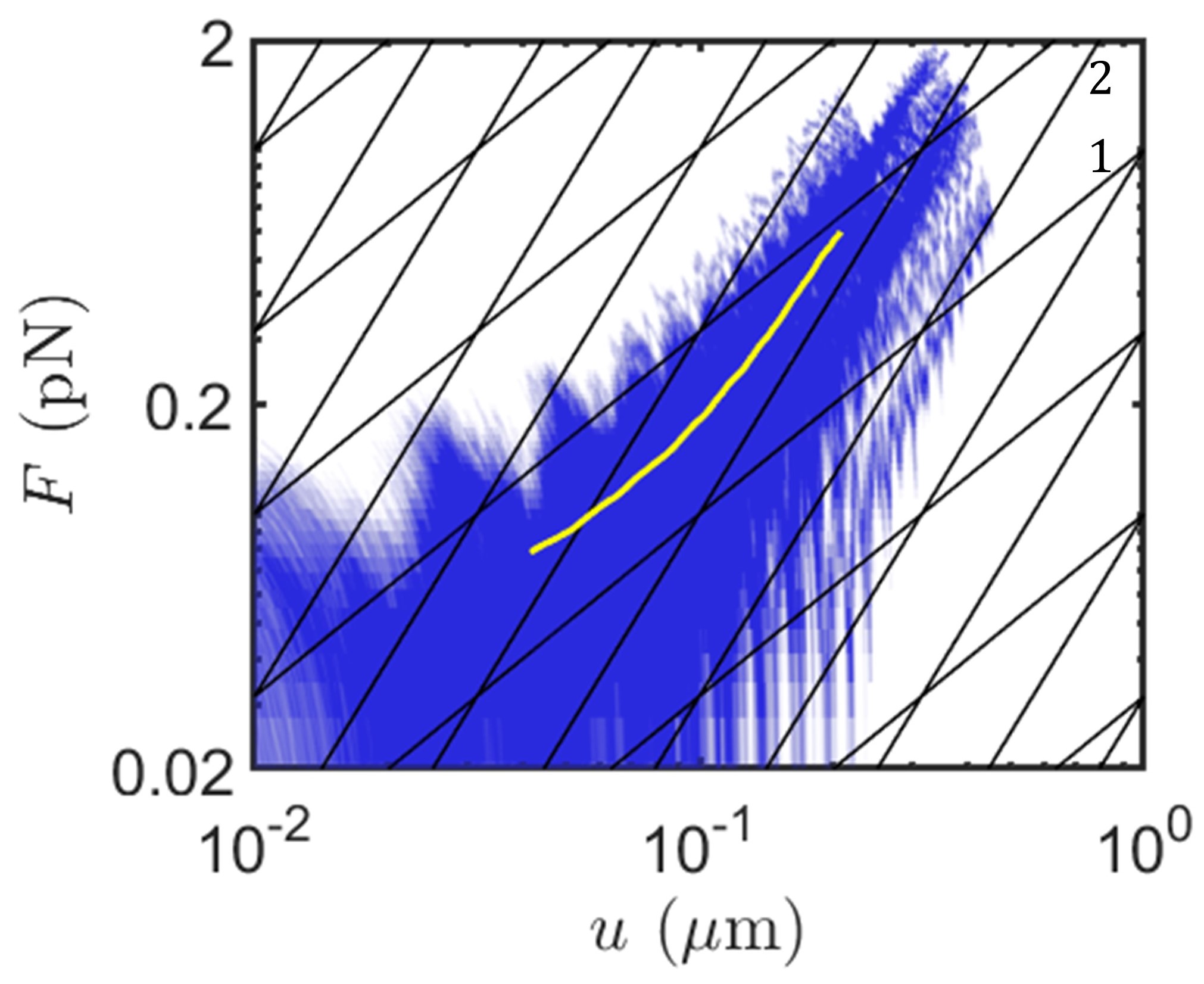}
\caption{For local measurements of collagen gels, the logarithmic slope of force as a function of displacement approaches 2 in the nonlinear regime, indicating force approaches a quadratic function of displacement. The yellow line shows the average force as a function of average displacement.}
\label{fig:FIGS1}
\end{figure}

\clearpage
\begin{figure}[H]
\centering
\includegraphics[width=0.48\textwidth]{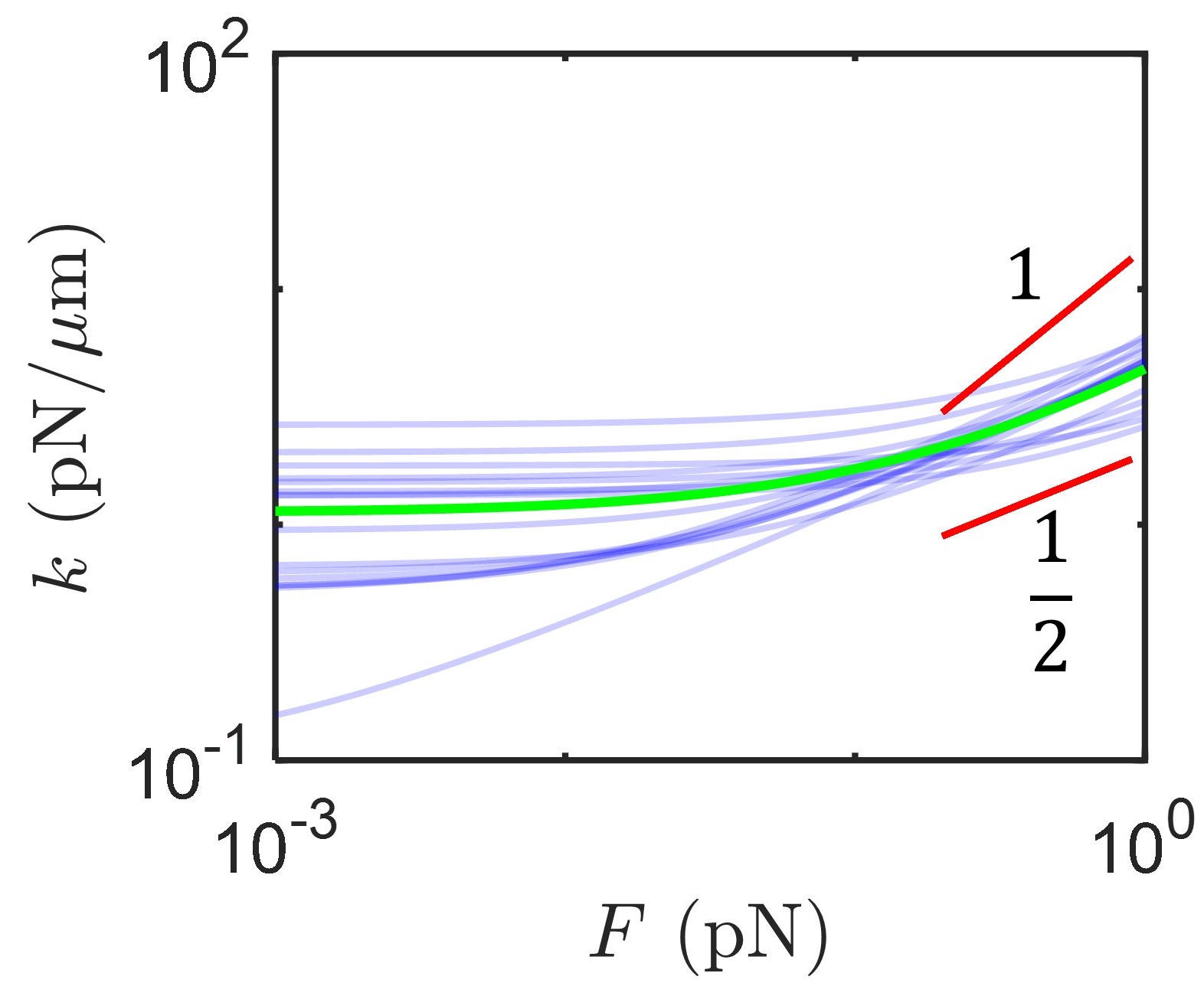}
\caption{Optical-tweezers measurements in \SI{3}{\mg \ml\tothe{-1}} fibrin gels (n=16). The green line is the average stiffness.}
\label{fig:FIGS2}
\end{figure}

\clearpage
\begin{figure}[H]
\centering
\includegraphics[width=0.48\textwidth]{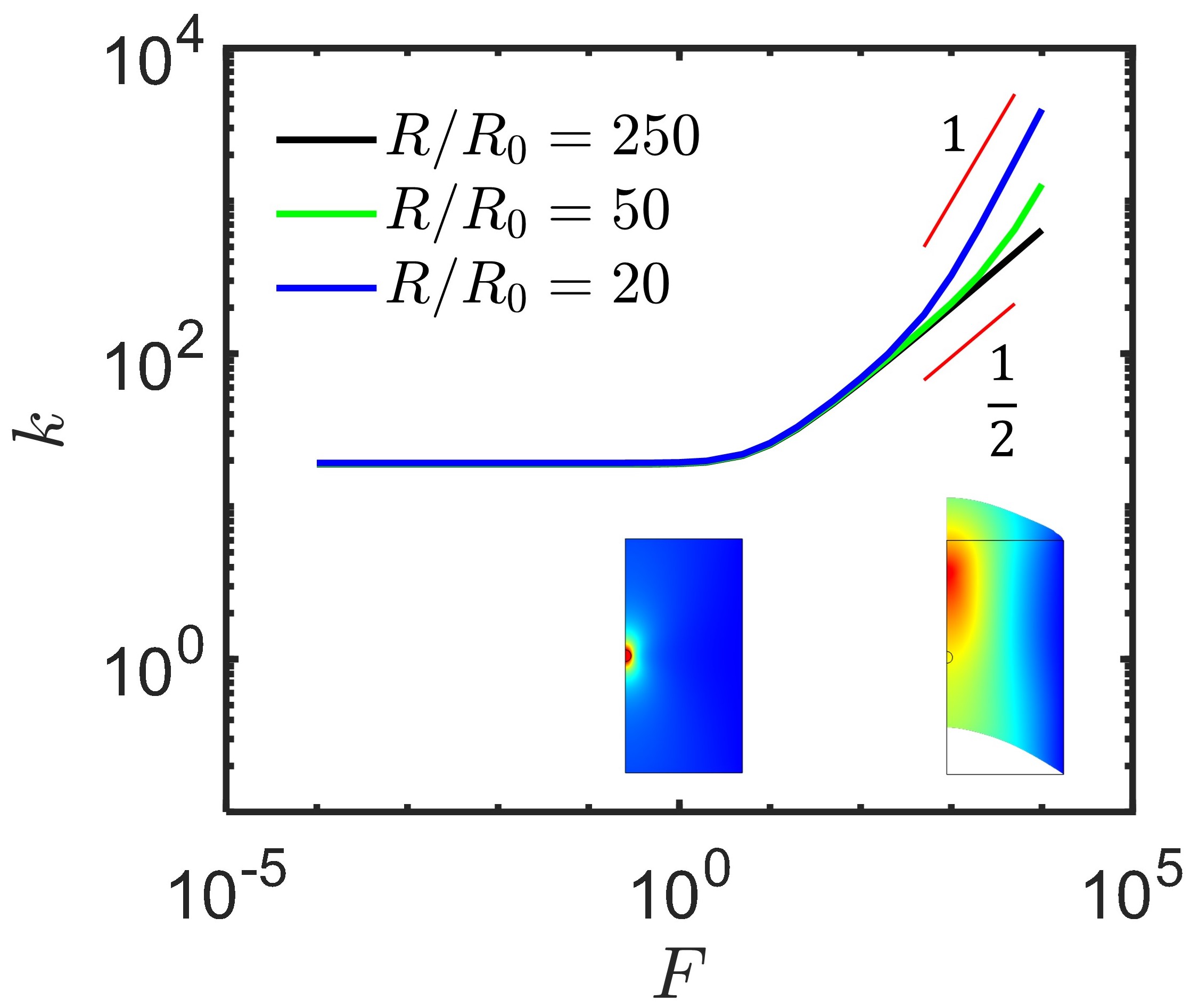}
\caption{Finite size effect. (Inset) Deformation of the sample $l/d=20$ is localized at small force but shows a global shearing deformation at large force. The contour color reflects normalized axial displacement.}
\label{fig:FIGS3}
\end{figure}

\clearpage
\begin{figure}[H]
\centering
\includegraphics[width=0.8\textwidth]{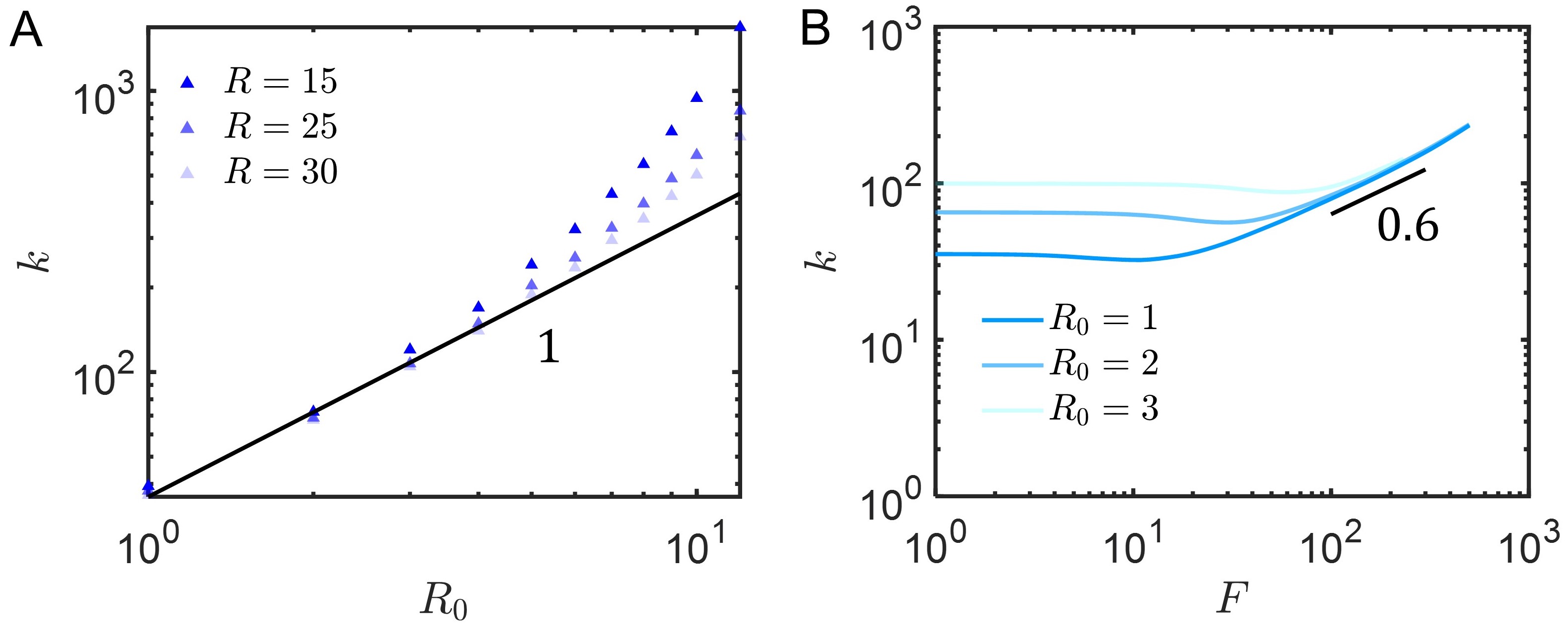}
\caption{Fiber network simulation. (\textit{A}) With increasing system size $R$, $k$ as a function of the particle size $R_0$ approaches a linear relation $k \sim R_0$. (\textit{B}) For various probe size $R_0$, $k \sim F^{0.6}$ in the nonlinear regime.}
\label{fig:FIGS4}
\end{figure}

\clearpage
\begin{figure}[H]
\centering
\includegraphics[width=0.48\textwidth]{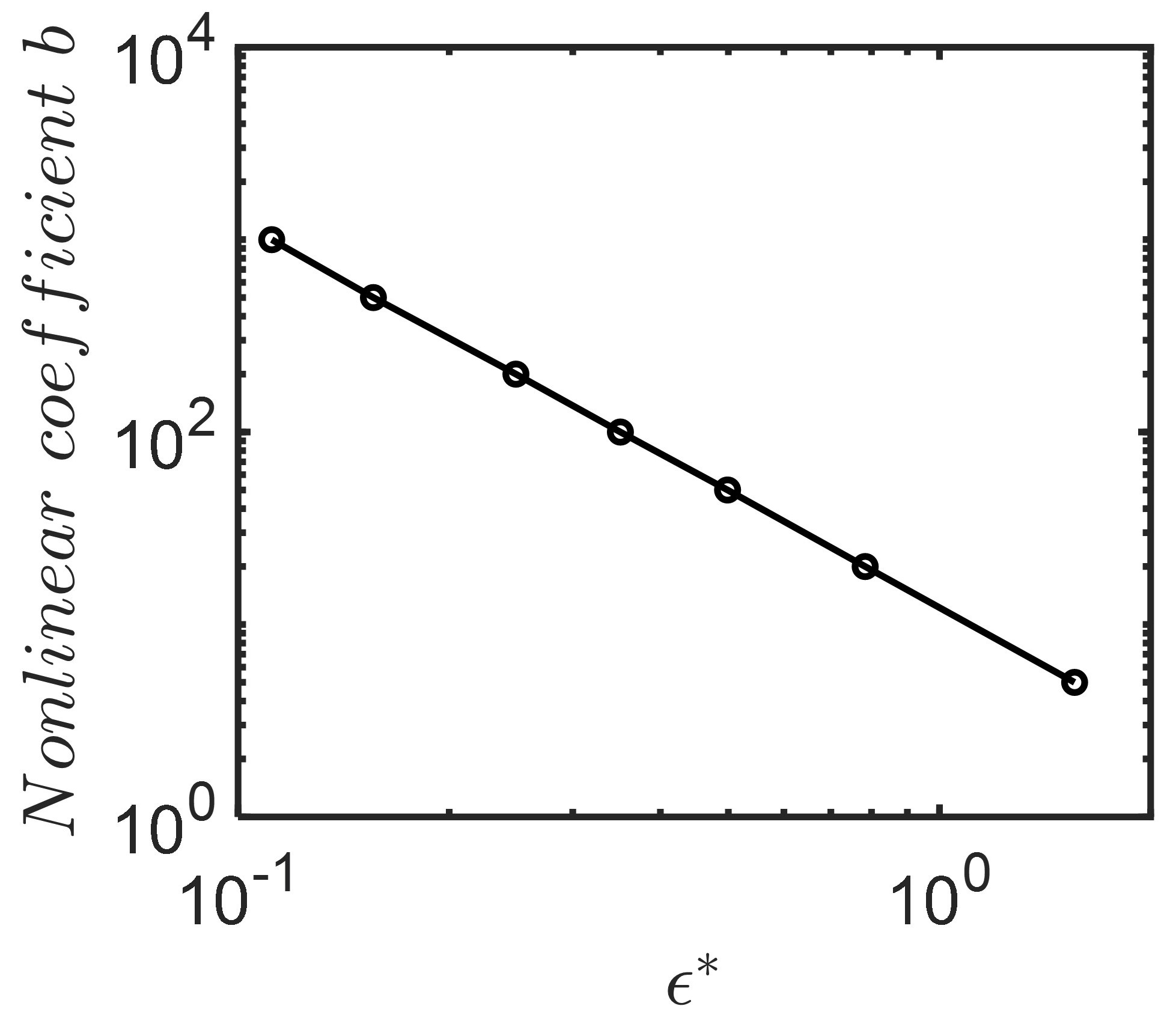}
\caption{The strain at the onset of stiffening may be used to decouple material nonlinearity from geometrical nonlinearity. Simulation shows the nonlinear stiffening parameter $b$ scales with the stiffening onset nominal strain $\epsilon ^*=2u^*/d$, where $u^*$ is the transition displacement $u^*=F^*/k^*$, and $d$ is the diameter of the bead.}
\label{fig:FIGS5}
\end{figure}

\end{document}